\documentclass[aps,prc,preprint,amsmath,amssymb,showpacs,preprintnumbers,superscriptaddress]{revtex4-1}
\usepackage{CJK}
\usepackage{graphicx}
\usepackage{dcolumn}
\usepackage{bm}
\usepackage{color}
\usepackage{hyperref}

\allowdisplaybreaks

\begin{document}
\begin{CJK*}{UTF8}{}
\title{Fission dynamics, dissipation and clustering at finite temperature}
\CJKfamily{gbsn}
\author{B. Li}
\affiliation{State Key Laboratory of Nuclear Physics and Technology, School of Physics, Peking University, Beijing 100871, China}
\author{D. Vretenar}
\email{vretenar@phy.hr}
\affiliation{Physics Department, Faculty of Science, University of Zagreb, 10000 Zagreb, Croatia}
\affiliation{State Key Laboratory of Nuclear Physics and Technology, School of Physics, Peking University, Beijing 100871, China}
\author{Z. X. Ren}
\affiliation{State Key Laboratory of Nuclear Physics and Technology, School of Physics, Peking University, Beijing 100871, China}
\author{T. Nik\v si\' c}
\affiliation{Physics Department, Faculty of Science, University of Zagreb, 10000 Zagreb, Croatia}
\affiliation{State Key Laboratory of Nuclear Physics and Technology, School of Physics, Peking University, Beijing 100871, China}
\author{J. Zhao}
\affiliation{Center for Circuits and Systems, Peng Cheng Laboratory, Shenzhen 518055, China}
\author{P. W. Zhao}
\email{pwzhao@pku.edu.cn}
\affiliation{State Key Laboratory of Nuclear Physics and Technology, School of Physics, Peking University, Beijing 100871, China}
\author{J. Meng}
\email{mengj@pku.edu.cn}
\affiliation{State Key Laboratory of Nuclear Physics and Technology, School of Physics, Peking University, Beijing 100871, China}

\begin{abstract}
The saddle-to-scission dynamics of the induced fission process is explored using a microscopic finite-temperature model based on time-dependent nuclear density functional theory (TDDFT), that allows to follow the evolution of local temperature along fission trajectories. Starting from a temperature that corresponds to the experimental excitation energy of the compound system, the model propagates the nucleons along isentropic paths toward scission. For the four illustrative cases of induced fission of $^{240}$Pu, $^{234}$U, $^{244}$Cm, and $^{250}$Cf, characteristic fission trajectories are considered, and the partition of the total energy into various kinetic and potential energy contributions at scission is analyzed, with special emphasis on the energy dissipated along the fission path and the prescission kinetic energy. The model is also applied to the dynamics of neck formation and rupture, characterized by the formation of few-nucleon clusters in the low-density region between the nascent fragments.
\end{abstract}

\date{\today}

\maketitle

\end{CJK*}
\section{Introduction}
Theoretical studies of induced nuclear fission dynamics have seen a strong revival in the last decade, prompted by a wealth of new experimental results and advances in microscopic methods that can be used to develop accurate models for large-scale calculation of fission observables \cite{schunck16,schmidt18,bender20,bulgac20,schunck22}.  Extensive studies of various aspects of the fission process have been reported, based on two principal microscopic approaches:
the time-dependent generator coordinate method (TDGCM) \cite{krappe12,schunck16,younes19,Regnier2016_PRC93-054611,Verriere2020_FP8-233} , and time-dependent density functional theory (TDDFT)  \cite{Ren_22PRL,simenel12,simenel18,nakatsukasa16,stevenson19,bulgac16,magierski17,scamps18,bulgac19,bulgac20}.
The former is a fully quantum mechanical approach, in which the nuclear wave function is represented by a superposition of generator states that are functions of collective coordinates. TDGCM can be applied to an adiabatic description of the entire fission process. It is especially suited to model the slow evolution from the quasi-stationary initial state to the outer fission barrier (saddle point) but, since only collective degrees of freedom are explicitly considered, this framework generally does not provide any dissipation mechanism. Various extension of the basic implementation of the TDGCM have been considered but, so far, no large-scale realistic calculation of dissipative fission dynamics has been reported \cite{bernard11,dietrich10,zhao22}. Beyond the outer fission barrier collective dynamics is coupled to intrinsic nucleon motion, and the resulting dissipative dynamics is usually modeled by TDDFT-based methods. Since TDDFT describes the classical evolution of independent nucleons in  mean-field potentials, it cannot be applied in the classically forbidden region of the collective space nor does it take into account quantum fluctuations.

Most microscopic studies have so far been focused on low-energy induced fission dynamics. To model the dependence of fission observables on excitation energy, one has to explicitly take into account the temperature of the compound nuclear system in a microscopic framework. Over the years several models have been developed that consider fission dynamics at finite temperature, both in the TDGCM framework \cite{schunck15,Zhao2019_PRC99-014618,Zhao2019_PRC99-054613,Schunck2020_LLNL-PROC-767225,McDonnell2014_PRC90-021302R,McDonnell2013_PRC87-054327,Martin2009_IJMPE18-861}, as well as based on the TDDFT \cite{zhu16,qiang21,PhysRevC.104.054604}. However, so far these models have not explicitly considered local changes in nuclear temperature and, therefore, cannot  describe the evolution of temperature as the fissioning nucleus evolves toward scission.

In this work we develop a TDDFT-based microscopic finite-temperature method, that allows to model the evolution of temperature along fission trajectories. Starting from a temperature that corresponds to the experimental excitation energy of the compound system, the model propagates the nucleons toward scission and beyond. At each step during the time evolution, the local temperature is adjusted so that the total energy is conserved. The present implementation of the model does not include the dynamical treatment of pairing correlations at finite temperature and, thus, can only be applied to cases in which pairing correlations essentially vanish. The theoretical framework, both at zero and finite temperature, is outlined in Sec.~\ref{sec_theo}. The dissipative saddle-to-scission dynamics, for the illustrative cases of induced fission of $^{240}$Pu, $^{234}$U, $^{244}$Cm, and $^{250}$Cf, is explored in Sec.~\ref{sec:paths}. Section \ref{sec:clusters}  includes an application to the dynamics of neck formation and rupture, determined by the formation of few-nucleon clusters in the low-density region between the emerging fission fragments. Finally, the principal results are summarized in Sec.~\ref{sec_summ}.

\section{Theoretical framework: TDDFT with explicit temperature dependence}\label{sec_theo}

The dissipative dynamics of the saddle-to-scission phase of the fission process will be modeled with the time-dependent covariant DFT~\cite{ren20LCS, ren20O16}. At zero temperature,
pairing correlations are treated dynamically with the time-dependent BCS approximation~\cite{ebata10TDBCS, Scamps13TDBCS}.
The wave function of the system takes the general form of a quasiparticle vacuum,
\begin{equation}
   |\Psi(t)\rangle = \prod_{k>0}\left[u_k(t)+v_k(t)c_k^+(t)c_{\bar{k}}^+(t)\right]|0\rangle,
\end{equation}
where $u_k(t)$ and $v_k(t)$ are the parameters in the transformation between the canonical and the quasiparticle states,
and $c_k^+(t)$ stands for the creation operator associated with the canonical state $\psi_k(\bm{r},t)$.
The evolution of $\psi_k(\bm{r},t)$ is determined by the time-dependent Dirac equation
\begin{equation}\label{Eq_td_Dirac_eq_BCS}
  i\frac{\partial}{\partial t}\psi_k(\bm{r},t)=\left[\hat{h}(\bm{r},t)-\varepsilon_k(t)\right]\psi_k(\bm{r},t),
\end{equation}
where the single-particle energy $\varepsilon_k(t)=\langle\psi_k|\hat{h}|\psi_k\rangle$, and the single-particle Hamiltonian $\hat{h}(\bm{r},t)$ reads
\begin{equation}
   \hat{h}(\bm{r},t) = \bm{\alpha}\cdot(\hat{\bm{p}}-\bm{V})+V^0+\beta(m_N+S).
   \label{Ham_D}
\end{equation}
The scalar $S(\bm{r},t)$ and four-vector $V^{\mu}(\bm{r},t)$ potentials are consistently determined at each step in time by the time-dependent densities and currents in the isoscalar-scalar, isoscalar-vector and isovector-vector channels,
\begin{subequations}\label{Eq_density_current}
  \begin{align}
    &\rho_S(\bm{r},t)=\sum_kn_k\bar{\psi}_k\psi_k,\\
    &j^\mu(\bm{r},t)=\sum_kn_k\bar{\psi}_k\gamma^\mu\psi_k,\\
    &j_{TV}^\mu(\bm{r},t)=\sum_kn_k\bar{\psi}_k\gamma^\mu\tau_3\psi_k,
  \end{align}
\end{subequations}
respectively. $\tau_3$ is the isospin Pauli matrix. The time evolution of the
occupation probability $n_k(t)=|v_k(t)|^2$, and pairing tensor $\kappa_k(t)=u_k^*(t)v_k(t)$, is governed by the following equations
\begin{subequations}\label{Eq_td_nkapp_eq_BCS}
   \begin{align}
     &i\frac{d}{dt}n_k(t)=n_k(t)\Delta_k^*(t)-n_k^*(t)\Delta_k(t),\\
     &i\frac{d}{dt}\kappa_k(t)=[\varepsilon_k(t)+\varepsilon_{\bar{k}}(t)]\kappa_k(t)+\Delta_k(t)[2n_k(t)-1] ,
   \end{align}
\end{subequations}
(for details, see Ref.~\cite{Scamps13TDBCS,ebata10TDBCS}). In time-dependent calculations, a monopole pairing interaction is employed, and the gap parameter $\Delta_k(t)$ is determined by the single-particle energy and pairing tensor,
\begin{equation}
  \Delta_k(t)=\left[G\sum_{k'>0}f(\varepsilon_{k'})\kappa_{k'}\right]f(\varepsilon_k),
\end{equation}
where $f(\varepsilon_k)$ is the cut-off function for the pairing window \cite{Scamps13TDBCS}.

In calculations with time-dependent covariant DFT, the mesh spacing of the lattice is 1.0 fm for all directions, and the box size is $L_x\times L_y\times L_z=20\times20\times60~{\rm fm}^3$.
The time-dependent Dirac equation \eqref{Eq_td_Dirac_eq_BCS} is solved with the predictor-corrector method, and the time-dependent equations \eqref{Eq_td_nkapp_eq_BCS} using the Euler algorithm.
The step for the time evolution is $6.67\times10^{-4}$~zs. For the
particle-hole channel we employ the point-coupling relativistic energy density functional PC-PK1~\cite{zhao10}. The pairing strength parameters: $-0.135$ MeV for neutrons, and $-0.230$ MeV for protons, are determined by the empirical pairing gaps of $^{240}$Pu,
using the three-point odd-even mass formula~\cite{Bender2000pairing}.
The initial states for the time evolution are obtained by self-consistent deformation-constrained relativistic DFT calculations in a three-dimensional lattice space, using the inverse Hamiltonian and Fourier spectral methods~\cite{ren17dirac3d, ren19LCS, ren20_NPA}, with the box size: $L_x\times L_y\times L_z=20\times20\times50~{\rm fm}^3$.

If one assumes that at the initial time the compound nucleus is in a state of thermal equilibrium at temperature $T$,
the system can be described by the finite temperature (FT) Hartree-Fock-Bogoliubov (HFB) theory \cite{goodman1981finite}.
In the grand-canonical ensemble, the expectation value of any operator $\hat{O}$ is given by an ensemble average
\begin{equation}
\langle \hat{O} \rangle = \textrm{Tr} ~[ \hat{D}\hat{O} ],
\end{equation}
where $\hat{D}$ is the density operator:
\begin{equation}
\hat{D} = {1 \over \mathcal{Z} } ~ e^{ -\beta \left( \hat{H}-\lambda \hat{N} \right) }\; .
\end{equation}
$\mathcal{Z}$ is the grand partition function, $\beta=1/k_{B}T$ with the Boltzmann constant $k_{B}$, $\hat{H}$ is the Hamiltonian of the system,
$\lambda$ denotes the chemical potential, and $\hat{N}$ is the particle number operator.

In the examples that will be considered in the next section, the internal excitation energy $E^*_{FS}$ of the fissioning system, defined as the the difference between the total binding energy of the equilibrium self-consistent mean-field  minimum at temperature $T$ and at $T = 0$, corresponds to temperatures that are above the pairing phase transition. The  temperature at which pairing correlations vanish depends on a specific nucleus but, for induced fission of actinides considered in the present work, the pairing energy is negligible at temperatures $T \geq 0.6$ MeV. In that case the FT HFB theory reduces to the self-consistent FT Hartree-Fock equations:
\begin{equation}\label{Eq_Dirac}
    \hat{h}\psi_k(\bm{r}) = \varepsilon_k\psi_k(\bm{r}),
\end{equation}
where the Dirac Hamiltonian $\hat{h}$ Eq.~(\ref{Ham_D}) is associated with a variation of the relativistic density functional PC-PK1~\cite{zhao10}:
\begin{equation}\label{Eq_energy_functional}
  \begin{split}
    E_{\rm tot}=\,&E_{\rm kin}+E_{\rm int}+E_{\rm em}\\
    =\,&\int d^3r~\left\{\sum_{k=1}^A\psi_k^\dag(\bm{\alpha}\cdot\hat{\bm{p}}+\beta m_N)\psi_k+\frac{1}{2}\alpha_S\rho_S^2+\frac{1}{3}\beta_S\rho_S^3+\frac{1}{4}\gamma_S\rho_S^4+\frac{1}{2}\delta_S\rho_S\Delta\rho_S\right.\\
    &+\frac{1}{2}\alpha_Vj^\mu j_\mu+\frac{1}{4}\gamma_V(j^\mu j_\mu)^2+\frac{1}{2}\delta_V j^\mu\Delta j_\mu+\frac{1}{2}\alpha_{TV}j^\mu_{TV}(j_{TV})_\mu+\frac{1}{2}\delta_{TV}j_{TV}^\mu\Delta(j_{TV})_\mu\\
    &+\left.ej_c^\mu A_\mu+\frac{1}{2}A_\mu\Delta A^\mu\right\},
  \end{split}
\end{equation}
and the scalar $S(r)$ and vector fields $V^\mu(r)$ read:
\begin{subequations}
  \begin{align}
    S(\bm{r})=\,&\alpha_S\rho_S+\beta_S\rho_S^2+\gamma_S\rho_S^3+\delta_S\Delta\rho_S,\\
    V^\mu(\bm{r})=\,&\alpha_Vj^\mu+\gamma_V(j^\mu j_\mu)j^\mu+\delta_V\Delta j^\mu+\tau_3\alpha_{TV}j_{TV}^\mu+\tau_3\delta_{TV}\Delta j_{TV}^\mu+e\frac{1-\tau_3}{2}A^\mu.
  \end{align}
\end{subequations}
In the absence of pairing correlations at finite temperature $T$, the local densities and currents $\rho_S$, $j^\mu$, and $j_{TV}^\mu$ can be written in the following form:
\begin{subequations}\label{Eq_density_current}
  \begin{align}
    &\rho_S=\sum_{k=1}^Af_k\bar{\psi}_k\psi_k,\\
    &j^\mu=\sum_{k=1}^Af_k\bar{\psi}_k\gamma^\mu\psi_k,\\
    &j_{TV}^\mu=\sum_{k=1}^Af_k\bar{\psi}_k\gamma_\mu\tau_3\psi_k,
  \end{align}
\end{subequations}
where $f_k$ is the thermal occupation probability,
defined as a function of single-particle energy $\varepsilon_k$ in Eq.~\eqref{Eq_Dirac}, the temperature $T$, and chemical potential $\lambda$:
\begin{equation}
  f_k = \frac{1}{1+e^{(\varepsilon_k-\lambda)/k_BT}}.
\label{thermal_f}
  \end{equation}
The chemical potential $\lambda$ is determined numerically in such a way that the particle number condition $\sum_k f_k = N$ is fulfilled.

In the dynamical case, the evolution of single-nucleon spinors $\psi_k$ is governed by the time-dependent Kohn-Sham equation~\cite{Rung1984TDDFT,Leeuwen1999TDDFT},
\begin{equation}\label{Eq_td_Dirac_eq}
  i\frac{\partial}{\partial t}\psi_k(\bm{r},t)=\hat{h}(\bm{r},t)\psi_k(\bm{r},t).
\end{equation}
The dependence on time of the Dirac Hamiltonian $\hat{h}(\bm{r},t)$ is determined by the time-dependent densities and currents~\cite{Rung1984TDDFT}.
The functional dependence of local densities and currents on temperature is the same as in the static case, with the time-dependent thermal occupation $f_k$,
\begin{equation}\label{eq_fkt}
  f_k(t) = \frac{1}{1+e^{[\varepsilon_k(t)-\lambda(t)]/k_BT(t)}}.
\end{equation}

The single-particle energy $\varepsilon_k(t)$ is defined: $\varepsilon_k(t)=\langle\psi_k(\bm{r},t)|\hat{h}(\bm{r},t)|\psi_k(\bm{r},t)\rangle$.
Note that in this case both $T(t)$ and $\lambda(t)$ are time-dependent.
Starting from the initial stationary values, the Lagrange multipliers $\lambda(t)$ and $T(t)$, considered as a non-equlibrium generalization of the chemical potential and temperature, are adjusted at each step in time in such a way that the particle number and total energy, respectively, are conserved along a TDDFT trajectory.

\section{Fission paths and energy dissipation}\label{sec:paths}

\begin{figure}[]
\centering
\includegraphics[width=1.0\textwidth]{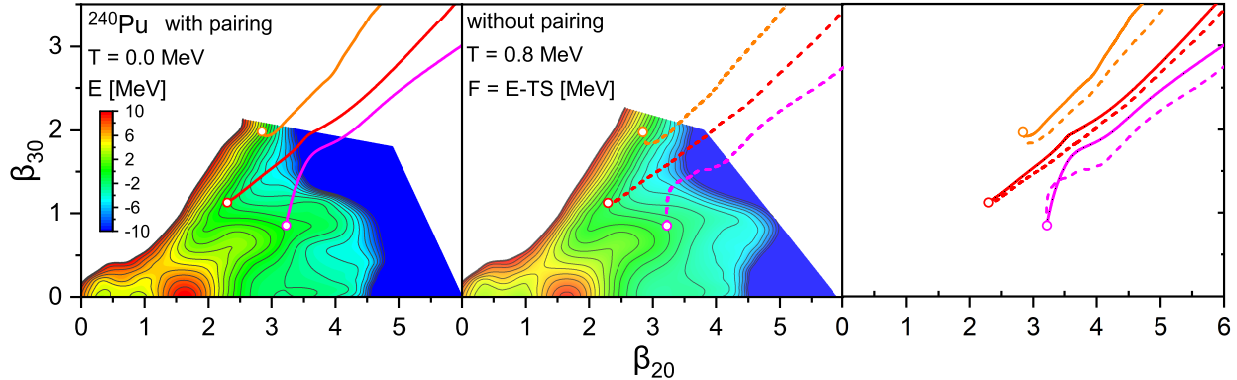}
\caption{Left panel: Self-consistent deformation energy surface of $^{240}$Pu in the plane of quadrupole-octupole axially-symmetric deformation parameters, calculated with the relativistic density functional PC-PK1 and a monopole pairing interaction at temperature $T=0$. Contours join points on the surface with the same energy (in MeV). The curves denote the TDDFT fission trajectories for three arbitrary initial points on the energy surface, located $\approx 1$ MeV below the energy of the equilibrium minimum. Middle panel: The corresponding self-consistent surface of Helmholtz free energy $F=E(T)-TS$, evaluated at the constant temperature $T = 0.8$ MeV. The three finite-temperature fission paths start at the same deformations like the $T=0$ paths in the left panel. Right panel: Comparison between $T=0$ and finite-temperature TDDFT fission paths.}
 \label{fig:240Pu}
\end{figure}

The panel on the left of Fig.~\ref{fig:240Pu} displays the self-consistent deformation energy surface of $^{240}$Pu, as function of the two collective coordinates: the axial quadrupole ($\beta_{20}$) and octupole ($\beta_{30}$) deformation parameters. As explained in the previous section, it is calculated using the relativistic energy density functional PC-PK1 and the monopole pairing interaction. The equilibrium minimum is located at $\beta_{20}\approx 0.3$ and $\beta_{30}=0$, the isomeric minimum is at $\beta_{20}\approx 0.9$ and $\beta_{30}=0$, and one notices the two fission barriers, and the fission valley at large deformations. The open dots denote three arbitrary initial points on the energy surface for calculation of fission trajectories. The TDDFT cannot be used to model the slow evolution from the equilibrium deformation to the saddle point \cite{bender20,schunck16,bulgac20,simenel14} and, therefore, the starting point is usually taken beyond the outer barrier \cite{scamps18,bulgac19}. The three points shown in the left panel of Fig.~\ref{fig:240Pu} correspond to energies approximately 1 MeV below the equilibrium minimum. Given the initial single-nucleon quasiparticle wave functions and occupation probabilities, TDDFT models a single fission events by propagating the nucleons independently toward scission and beyond. At each step in time the single-nucleon potentials are determined from the time-dependent densities, currents and pairing tensor and, thus, the time-evolution includes the one-body dissipation mechanism.

The three trajectories in the left panel are among those that we considered in two recent studies of fission dynamics. In Ref.~\cite{Ren_22PRC} low-energy induced fission of $^{240}$Pu has been analyzed using a consistent microscopic framework that combines the TDGCM and TDDFT. The former presents a fully quantum mechanical approach that describes the entire fission process as an adiabatic evolution of collective degrees of freedom, while the latter models the dissipative dynamics of the final stage of fission by the self-consistent time-evolution of single-nucleon wave functions toward scission. The study has shown that quantum fluctuations, included in TDGCM but not in TDDFT, are essential for a quantitative estimate of fission yields. Dissipative effects, taken into account in TDDFT but not in TDGCM, are crucial for the total kinetic energy distribution.

In Ref.~\cite{Ren_22PRL} TDDFT has been employed to study the dynamics of neck formation and rupture in the process of induced nuclear fission. By following mass-asymmetric fission trajectories in $^{240}$Pu, it has been shown that the time-scale of neck formation coincides with the assembly of two $\alpha$-like clusters ($\approx 100 - 200$ fm/c). The low-density region between the nascent fragments provides the conditions for dynamical synthesis of $^{4}$He and other light clusters. The neck ruptures at a point exactly between the two $\alpha$-like clusters, which separate because of the Coulomb repulsion and are eventually absorbed by the two emerging fragments.

In the present work we extend these studies to a more realistic description of induced fission dynamics that includes the effect of finite temperature of the compound nucleus. As we have already shown in the TDGCM with Gaussian overlap approximation (GOA) studies of mass-asymmetric fission of actinides in Refs.~\cite{Zhao2019_PRC99-014618} and \cite{Zhao2019_PRC99-054613}, the extension to finite temperature leads to a considerable improvement of the
calculated charge yields. The most serious limitation of the TDGCM+GOA approach is, of course, the fact that it does not include dissipation and the fissioning systems evolves toward scission at a constant temperature. To describe energy dissipation and heating of the nucleus as it evolves toward scission, in this study we apply the finite temperature extension of the TDDFT.

The TDGCM+GOA calculation of induced fission of $^{240}$Pu in Ref.~\cite{Zhao2019_PRC99-054613} was carried out at the constant temperature $T=0.8$ MeV, which corresponds to an average experimental excitation energy of $10.7$ MeV \cite{Ramos2018_PRC97-054612}. At this temperature pairing correlations vanish, and the thermodynamical potential relevant for the analysis of finite-temperature deformation effects is the Helmholtz free energy $F=E(T)-TS$, where the
entropy of the compound nuclear system is computed using the relation:
\begin{equation}
S = -k_{B} \sum_{k} \left[ f_{k} \ln f_{k} + (1 - f_{k}) \ln (1 - f_{k}) \right],
\label{eq:entropy}
\end{equation}
where $f_k$ is the thermal occupation function of Eq.~(\ref{eq_fkt}).
In the middle panel of Fig.~\ref{fig:240Pu} we plot the Helmholtz free energy $F=E(T)-TS$, evaluated at temperature $T = 0.8$ MeV. This is the initial temperature for the TDDFT evolution, and we will consider the three finite-temperature fission paths that start at the same deformations like the $T=0$ paths in the left panel. The panel on the right emphasizes the differences between the $T=0$ and finite-temperature TDDFT fission paths. It is interesting that, even though at $T=0.8$ MeV the dynamics is no longer determined by pairing correlations, the paths are not much different from the $T=0$ fission trajectories. The general effect of increasing the internal excitation energy, that is, the initial nuclear temperature, is to shift fission to more symmetric configurations of the resulting fragments.

Note that the assignment of the initial temperature to an arbitrary point on the energy surface is not entirely correct, as this temperature strictly corresponds to the compound nucleus at equilibrium deformation. However, it is generally accepted that dissipation between equilibrium and the outer barrier is weak, and only beyond the saddle point fission dynamics becomes strongly dissipative as the nucleus quickly elongates toward scission. Since, in any case, TDDFT cannot be used to model the equilibrium to outer barrier dynamics, it seems reasonable to assign the temperature of the compound nucleus to an initial point beyond the outer barrier. The actual value of the initial temperature is not that important, as it corresponds to an average excitation energy of the fissioning system. More interesting is the rate of change of local temperature along a fission path.

\begin{figure}[tbh!]
\centering
\includegraphics[width=0.65\textwidth]{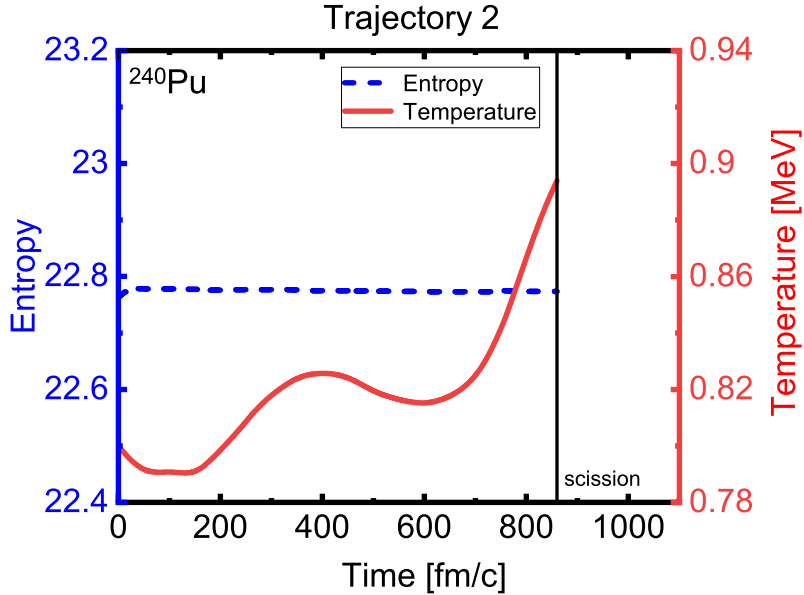}\\
 \caption{Local temperature and entropy as functions of time, for trajectory 2 shown in the middle panel of Fig.~\ref{fig:240Pu}.}
 \label{fig:S_T}
\end{figure}

For the illustrative case of trajectory 2 in the middle panel of Fig.~\ref{fig:240Pu}, in Fig.~\ref{fig:S_T} we plot the evolution in time of the local temperature and entropy, from the initial point to scission. TDDFT, of course, propagates the nucleon wave functions also beyond scission, however the resulting fission fragments will generally have different temperatures. This particular feature cannot be described in the present implementation of TDDFT, and this is why we only consider fission paths up to scission. We notice that, as one would expect for dissipative dynamics, the local temperature generally increases along the fission path. In this particular case, the temperature at scission is $T=0.89$ MeV, that is, the increase from the initial point is approximately ten percent. Other examples will be discussed further below. The local entropy calculated with Eq.~(\ref{eq:entropy}), on the other hand, remains constant along the fission path. This means that, even without any constraint on the entropy, our temperature-dependent TDDFT model describes an isentropic process of self-consistent evolution of the fissioning system.

\begin{figure}[]
\centering
\includegraphics[width=1.0\textwidth]{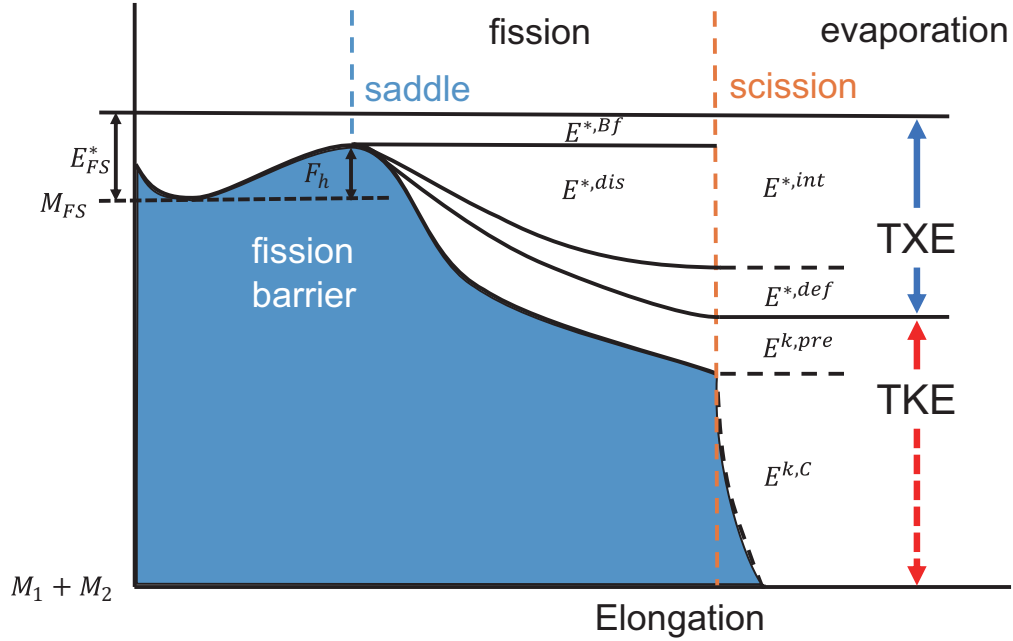}
\caption{Definition of the various components of the total energy of a nucleus along a typical fission path. See text for explanation. Adapted from Fig. 1 of Ref.~\cite{Caamano_17PLB}.}
 \label{fig:Energy_definitions}
\end{figure}

To discuss energy dissipation and heating along a typical fission path (cf. Table \ref{tab:dissipation}), in Fig.~\ref{fig:Energy_definitions}, which is adapted from Fig. 1 of Ref.~\cite{Caamano_17PLB}, we summarize the various components of the total energy as functions of the nuclear elongation. $M_{FS}$ is the mass of the fissioning system,  $E^{*}_{FS}$ is the average excitation energy, and the masses of the two fragments are $M_1$ and $M_2$. Then, assuming that there is no evaporation from saddle to scission (i.e. the fissioning nucleus remains a closed system), the energy balance can be expressed with the following relation \cite{Caamano_17PLB}
\begin{equation}
E^{*}_{FS} + M_{FS} = M_1 + M_2 +TKE +TXE
\label{eq:en_balance}
\end{equation}
The total kinetic energy TKE consists of the Coulomb energy $E^{k,C}$ between the fragments at scission, and the prescission kinetic energy $E^{k,pre}$ which results from a partial conversion of the saddle-to-scission collective potential energy difference (the other part is converted into the deformation energy of the fragments and dissipation energy). $E^{k,pre}$ is defined as the collective flow energy at scission \cite{bulgac19}
\begin{equation}
E^{k,pre} = \frac{m}{2} \int   \rho(\vec{r}, t_{sci}) {\vec{v}}~^2 (\vec{r},t_{sci}) d\vec{r}\;,
\label{eq:en_flow}
\end{equation}
where the density and velocity field are evaluated at the time of scission.
The total excitation energy TXE is divided into the deformation energy of the fragments at scission and the total intrinsic excitation energy,
\begin{equation}
TXE = \sum_{i=1}^2 E_i^{*,def} + E^{*,int}\; .
\label{eq:en_TXE}
\end{equation}
The former can be easily computed by taking, for each fragment, the difference between the $T=0$ deformation-constrained energy of the fragment at scission and its mass (energy at equilibrium deformation). The expression for the total intrinsic excitation energy $E^{*,int}$ reads:
\begin{equation}
E^{*,int} = E^{*,Bf} + E^{*,dis}\; ,
\label{eq:en_E^int}
\end{equation}
where $E^{*,Bf}$  is the difference between the total energy of the nucleus and the energy at the saddle point (see Fig.~\ref{fig:Energy_definitions}), and $E^{*,dis}$ is the energy dissipated along the fission path. The partition of the total intrinsic excitation energy between the fragments can be calculated under additional model assumptions  \cite{Caamano_17PLB}, but here this is not crucial as we only follow the dynamics up to scission.

The results for fission trajectories 2 and 3 of $^{240}$Pu, shown in the middle panel of Fig.~\ref{fig:240Pu}, are listed in the first two columns of Table \ref{tab:dissipation}, respectively. The first two lines include the temperature ($T=0.8$ MeV) and total energies at the initial point. This energy is, as explained in the previous section, fully conserved along the fission path. In the next two lines we list the prescission kinetic energies $E^{k,pre}$ (4.33 MeV and 5.45 MeV for paths 2 and 3, respectively) and Coulomb energies between the fragments at scission $E^{k,C}$ (180.32 MeV and 169.01 MeV for paths 2 and 3, respectively). The sum $E^{k,pre} + E^{k,C}$ is the total kinetic energy.  The next four lines contain, for each fission fragment, the ground state energy and deformation energy at scission. $E^{*,int}$ is the total intrinsic excitation energy at scission (18.93 MeV and 27.27 MeV for trajectories 2 and 3, respectively), $E^{*}_{FS}$ is the excitation energy that corresponds to the initial temperature, and $F_h$ is the height of the fission barrier at the initial temperature. $E^{*,Bf}$ is the available energy above the saddle point, $E^{*,dis}$ is the dissipation energy and, finally, $T_{sci}$ is the temperature at the scission point. For trajectory 2 the dissipated energy at scission is 12.64 MeV, and the corresponding increase in temperature is 0.09 MeV. For trajectory 3 these values are: $E^{*,dis} = 20.98$ MeV and $\Delta T = 0.2$ MeV. The results for the very asymmetric trajectory 1 are not included because of numerical problems in obtaining convergence in the constrained calculation of the deformation energy of the lighter fragment.

\begin{figure}
\centering
\includegraphics[width=1.0\textwidth]{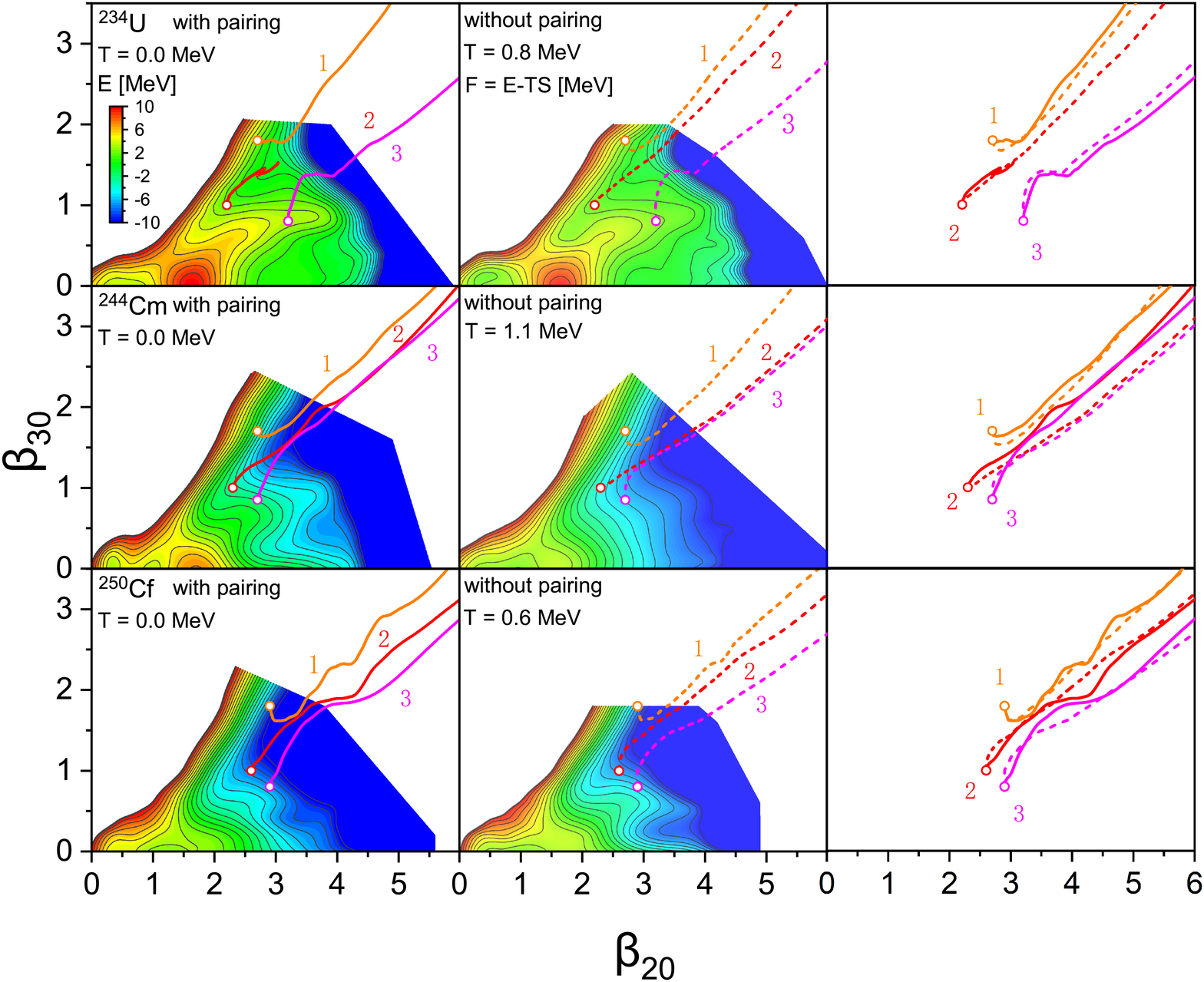}
 \caption{Same as in the caption to Fig.~\ref{fig:240Pu} but for the process of induced fission of $^{234}{\rm U}$~(top),~$^{244}{\rm Cm}$~(middle),~and~$^{250}{\rm Cf}$~(bottom).
  }
 \label{fig:UCmCf}
\end{figure}

In addition to $^{240}$Pu, we have computed similar fission paths for three more actinides that were also included in the finite-temperature TDGCM+GOA study of Ref.~\cite{Zhao2019_PRC99-054613}. For $^{234}$U the initial temperature $T=0.8$ MeV corresponds to the experimental peak photon energy $E_{\gamma} = 11$ Mev in photo-induced fission \cite{Schmidt_00NPA}. The temperature $T=1.1$ MeV, that we choose in the case of $^{244}$Cm, equates to an average experimental excitation energy of 23 MeV for multinucleon transfer-induced fission \cite{Ramos2018_PRC97-054612}. Finally, the initial temperature $T=0.6$ MeV of $^{250}$Cf corresponds to thermal neutron-induced fission \cite{Brown_18NDS}. Just like in the case of $^{240}$Pu, in Fig.~\ref{fig:UCmCf} for $^{234}$U, $^{244}$Cm, and $^{250}$Cf, we display the deformation energy surface at zero temperature, the Helmholtz free energy at finite initial temperature, and three characteristic fission paths that start from the same deformations at zero and finite initial $T$. In all four cases, the initial temperatures for the compound nuclei are above the pairing phase transition and, therefore, pairing correlations are not taken into account during the time evolution toward scission.

Considering the deformation energy surfaces, one notices that the fission barriers are significantly reduced at finite temperatures but, of course, for initial points beyond the saddle, the fission trajectories at $T=0$ and finite temperature are not very different. In general, the trajectories follow the path of steepest descent. An exception is the trajectory 2 for $^{234}$U which, in the case of zero temperature, remains confined in a region of a local minimum or saddle, and does not proceed to scission. This is a well known effect in TDDFT modeling of fission. As we have shown in the recent microscopic analysis of fission dynamics of $^{240}$Pu \cite{Ren_22PRC}, at zero temperature not all TDDFT trajectories that start below the outer barrier lead to scission and formation of fission fragments. The results for the final temperature, prescission kinetic energy, intrinsic excitation energy, and dissipated energy at scission, are consistent with those obtained for $^{240}$Pu (cf. Table \ref{tab:dissipation}). The increase in temperature from the initial points to scission is generally in the interval $10\% - 20\%$. The prescission kinetic energy is of the order of $4 - 9$ MeV, and this means that a relatively small portion of the potential energy difference at scission is converted into collective flow energy. In fact, as shown in the table, the dissipated energy $E^{*,dis}$ is at least a factor $2 - 4$ larger than $E^{k,pre}$, and so is the corresponding intrinsic excitation energy $E^{*,int}$. This result illustrates the importance of the one-body dissipation mechanism included in time-dependent nuclear density functional theory, in contrast to approaches that consider only collective degrees of freedom, such as the TDGCM+GOA. Finally, we note that, just as in the case of $^{240}$Pu, the most asymmetric fission paths (trajectory 1) in Fig.~\ref{fig:UCmCf}, lead to scission configurations for which it has not been possible to obtain fully converged solutions in  the constrained calculation of deformation energy of the fragments, and this is why the corresponding results are not included in the table.
\begin{table}
\caption{Initial temperature, total energy of the fissioning system at the initial point, various components of the total energy at scission, and the final temperature at scission, for trajectories 2 and 3 of $^{240}$Pu, $^{234}$U, $^{244}$Cm, and $^{250}$Cf, shown in Figs.~\ref{fig:240Pu} and \ref{fig:UCmCf}, respectively. All values are given in MeV.}
\bigskip
  \centering
  \renewcommand\arraystretch{1.5}
  \begin{tabular}{ccccccccc}
    \hline\hline
      Nucleus &\multicolumn{2}{c}{$^{240}{\rm Pu}$}&\multicolumn{2}{c}{$^{234}{\rm U}$}&\multicolumn{2}{c}{$^{244}{\rm Cm}$}&\multicolumn{2}{c}{$^{250}{\rm Cf}$}\\
    \hline
      Trajectory & 2 & 3 & 2 & 3 & 2 & 3 & 2 & 3 \\
     \hline
     $T_{init}$               &  0.80    &  0.80    & 0.80     & 0.80     &   1.10   &  1.10    &  0.60    &  0.60    \\

     $E_{tot}$                & -1801.15 & -1795.23  & -1757.51 & -1750.90 & -1812.95 & -1810.63 & -1859.50 & -1858.27 \\

     $E^{k,pre}$                &  4.33    &   5.45   & 5.12     & 5.13   &  5.36    &  7.50    & 6.52     &  9.13    \\

     $E^{k,C}$                  &  180.32  &   169.01 & 167.83   & 164.88 &  180.88  &  173.94  & 174.36   &  182.71  \\

     $E^1_{g.s.}$              & -1126.58 & -1101.47 & -1129.67 & -1073.62 & -1132.98 & -1134.96 & -1159.60 & -1143.63 \\

     $E^{*,def}_1$              & 3.40     & 11.08   & 3.34     & 10.10   & 6.78     & 6.77     & 3.39     & 3.34     \\

     $E^2_{g.s.}$               & -889.51  & -913.27  & -840.78  & -890.46 & -914.23  & -911.12  & -922.92  & -941.14  \\

     $E^{*,def}_2$               & 7.96     & 6.70  & 9.85     & 1.89     & 8.03     & 9.47     & 8.15     & 6.51     \\

     $E^{*,int}$                & 18.93     & 27.27 & 26.80      &31.18     &   33.21  &   37.77  &  30.60   &  24.81   \\

     $E^{*}_{FS}$               & 11.40    & 11.40  & 11.23    & 11.23    &   22.63  &   22.63  &  7.24    &  7.24    \\

     $F_h$                      & 5.11     &  5.11    & 5.46     & 5.46   &    3.02  &   3.02   &  4.06    &  4.06    \\

     $E^{*,Bf}$                 & 6.29     &  6.29  & 5.77     & 5.77     & 19.61    &   19.61  &  3.18    &  3.18    \\

     $E^{*,dis}$                & 12.64    &  20.98  & 21.03    & 25.41   & 13.60    &  18.16   &  27.42   & 21.63    \\

     $T_{sci}$                  &  0.89    &  1.00   & 0.85     & 0.97    &   1.20   &  1.27    &  0.72    &  0.73    \\

       \hline\hline
  \end{tabular}
\label{tab:dissipation}
\end{table}

\section{Clusters in the neck at scission}\label{sec:clusters}
A number of theoretical studies, starting with the pioneering work of Ref.~\cite{negele78}, have established the importance of including pairing correlations for computing spontaneous fission lifetimes and modeling induced fission observables. In particular, by employing various time-dependent approaches, it has been shown that the fission process can be retarded or even completely impeded by the exclusion of pairing, while an increase in strength of a pairing interaction leads to a significant acceleration of fission dynamics (cf. Refs.~\cite{bulgac19,bulgac20} and references therein). In a recent study based on the TDGCM+GOA \cite{zhao21}, we have analyzed the role of dynamical pairing in induced fission dynamics. A calculation of fragment charge yields, performed in a 3D space of collective coordinates that, in addition to the axial quadrupole and octupole intrinsic deformations, also includes an isoscalar pairing degree of freedom, has shown that the inclusion of dynamical pairing has a pronounced effect on the collective inertia, the collective flux through the scission hypersurface, and the resulting fission yields.

In many experimental situations, however, as also shown by the examples considered in the present study, the excitation energy of the compound system corresponds to a temperature well above the pairing phase transition. For the fission paths shown in Figs.~\ref{fig:240Pu} and \ref{fig:UCmCf}, in Table \ref{tab:time} we compare the time intervals from the initial point of a trajectory to the scission point. Except for trajectory number 2 of $^{234}$U which does not end up in scission at $T=0$, we do not find a significant difference in the time it takes to reach the scission point starting at zero temperature with pairing correlations included, or at finite temperatures at which pairing does not contribute to fission dynamics.

\begin{table}

  \centering
  \renewcommand\arraystretch{1.5}
  \caption{Time interval, in units of fm/c, from the initial point of a trajectory to the scission point.}
  \begin{tabular}{ccccccccccccccccc}
    \hline\hline
      Nucleus &&\multicolumn{3}{c}{$^{234}{\rm U}$}&&\multicolumn{3}{c}{$^{240}{\rm Pu}$}&&\multicolumn{3}{c}{$^{244}{\rm Cm}$}&&\multicolumn{3}{c}{$^{250}{\rm Cf}$}\\
    \hline
      Trajectory && 1 &2 & 3 && 1 & 2 & 3 && 1 & 2 & 3 && 1 & 2 & 3 \\

     \hline
     $T = 0$ w/ pairing~       && 960      & $-$ & 940    &&  1600    &  1150    &  700    &&  900    &  1080   &  820 &&  1040 &  900  &  600  \\

     $T \neq 0$ w/o pairing~ && 840      & 1100       & 1000   &&  1160    &  860     &  820    &&  880    &  1140   &  720 &&  1020 &  800  &  520  \\

  \hline\hline
  \end{tabular}
\label{tab:time}
\end{table}

Below saturation density, nuclear matter becomes inhomogeneous and, at low densities, the nucleus can locally minimize its energy by forming light clusters, in particular strongly bound $\alpha$-particles \cite{typel10,zinner13,ebran14,ebran17}.
Extensive experimental and/or theoretical studies of the formation of light clusters of nucleons has been performed in a variety of environments, such as light and medium-heavy $N=Z$ and neutron-rich nuclei~\cite{Ebran2012Nature341,Datar13,Ichikawa11,Zhao15}, the surface (skin) region of heavy nuclei~\cite{Xu2017,Tanaka21}, expanding hot matter in heavy-ion reactions~\cite{Petrovici1995_PRL74-5001}, and core-collapse supernovae~\cite{Sumiyoshi2008_PRC77-055804}. In the context of the present analysis, of particular interest is the formation of clusters in the low-density neck region of a fissioning nucleus \cite{wuenschel14,ropke21,denisov21,Ren_22PRL}, as manifested by the kinematics of ternary fission events in which not only $^{4}$He, but also $^{3}$H and $^{6}$He cluster emission is observed. In the recent TDDFT study of the final phase of the fission process that precedes scission \cite{Ren_22PRL}, we have shown that the mechanism of neck formation and its rupture are characterized by the dynamics of light clusters. In a mean-field analysis, however, one cannot directly identify few-nucleon clusters and, as shown in Ref.~\cite{Ren_22PRL}, the one-body density at the time of scission does not exhibit signatures of cluster formation.
One must rather consider the corresponding time-dependent nucleon localization functions \cite{becke90,reinhard11}:
\begin{equation} \label{nlf}
C_{q\sigma}(\vec{r})=\left[1+\left(\frac{\tau_{q\sigma}\rho_{q\sigma}-{1\over 4}|\vec{\nabla}\rho_{q\sigma}|^2-\vec{j}^2_{q\sigma}}{\rho_{q\sigma}\tau^\mathrm{TF}_{q\sigma}}\right)^2\right]^{-1} ,
\end{equation}
for the spin $\sigma$ ($\uparrow$ or $\downarrow$)  and isospin $q$ ($n$ or $p$) quantum
numbers. $\rho_{q\sigma}$, $\tau_{q\sigma}$, $\vec{j}_{q\sigma}$, and $\vec{\nabla}\rho_{q\sigma}$ denote the nucleon density, kinetic energy density, current density, and density gradient, respectively.
$\tau^\mathrm{TF}_{q\sigma}={3\over 5}(6\pi^2)^{2/3}\rho_{q\sigma}^{5/3}$ is the Thomas-Fermi kinetic energy density.

For homogeneous nuclear matter $\tau = \tau^\mathrm{TF}_{q\sigma}$, the second and third term in the numerator vanish,
and $C_{q\sigma} = 1/2$. In the other limit $C_{q\sigma} (\vec{r}) \approx 1$
indicates that the probability of finding two nucleons with the same spin and isospin at the same point $\vec{r}$ is very small.
This is the case for the $\alpha$-cluster of four particles: $p \uparrow$,  $p \downarrow$, $n \uparrow$,
and $n \downarrow$, for which all four nucleon localization functions $C_{q\sigma} \approx 1$.

For the illustrative case of induced fission of $^{240}$Pu \cite{Ren_22PRL}, a detailed analysis of several characteristic  trajectories has shown that, while the localization functions generally exhibit shell structures in the fissioning system and the fragments, their values $0.4$ -- $0.6$ are consistent with homogeneous nuclear matter. At times immediately preceding scission, however, values close to 1 are obtained in the neck region, characteristic for $\alpha$-clusters. The emergence of pronounced localization coincides with the formation of the neck between the two large fragments in a short time interval $\approx 100 - 200$ fm/c. The scission event then occurs between two $\alpha$-like clusters, which repel because of Coulomb interaction and are absorbed by the fragments, where they induce strongly damped dipole oscillations along the fission axis. Even though, by using the TDDFT mean-field method, one cannot uniquely identify the content of each cluster in the neck region, an integration of the one-body density showed that the elongation of the neck at scission corresponds to the region that contains, in total, four protons and approximately eight neutrons. The principal result is a new mechanism of neck rupture, determined by the formation of $\alpha$-like clusters. If, at the moment of scission, one of the clusters is not absorbed by the corresponding large fragment, it will be emitted perpendicular to the fission axis by the Coulomb repulsion with the fragments, resulting in a ternary fission event.

Only $^{240}$Pu was considered in the induced fission analysis of Ref.~\cite{Ren_22PRL} and, thus, to verify the validity of the proposed mechanism of cluster formation in the low-density neck region and the subsequent scission event, here we examine two more cases: $^{250}$Cf and $^{244}$Cm. The reason for this specific choice is that we also want to analyze the effect of increasing temperature along a fission trajectory on the formation of clusters in the neck region. Temperature increase was not considered in our previous study, and this has been one of the reasons for developing a finite-temperature TDDFT formalism that can be used to describe the effect of heating dilute nuclear matter in the region where scission occurs. In general, one expects that localization and cluster formation are suppressed when the temperature of nuclear matter increases. In a very recent relativistic Hartree-Bogoliubov study of clustering effects in $^{20}$Ne and $^{32}$Ne at finite temperature \cite{yuksel22}, it has been shown that clustering features gradually weaken with increasing temperature, and disappear as the shape of the nucleus changes from prolate to spherical. The pronounced equilibrium prolate deformation in these nuclei is strongly reduced with increasing temperature and, in fact, a shape phase transition is observed at the mean-field level, leading to a complete dissolution of $\alpha$-like clusters. In the present case the situation is somewhat different because, as the temperature increases, the elongation of the fissioning system increases and a low-density neck region between the fragments appears.

\begin{figure}[tbh!]
\centering
\includegraphics[width=0.75\textwidth]{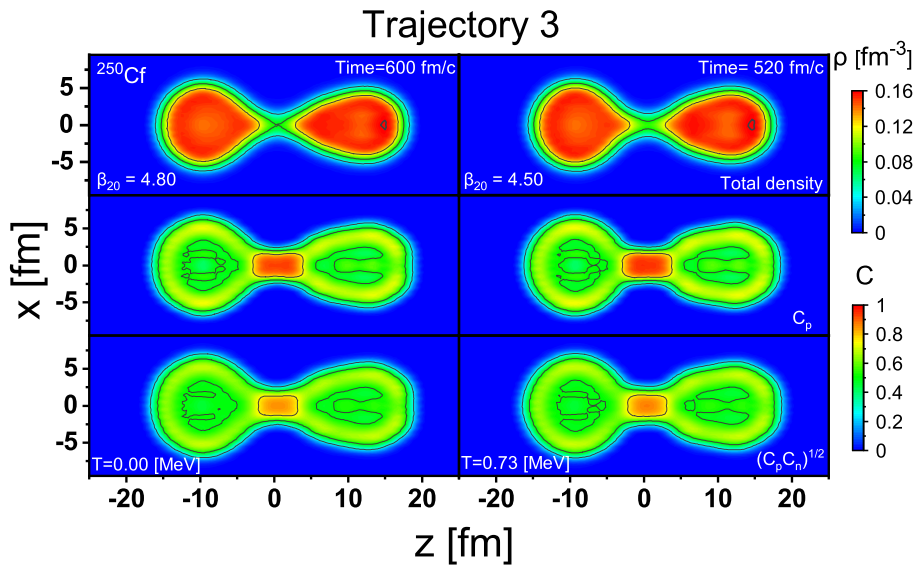}
\caption{Left top: density profile of $^{250}$Cf (color code in fm$^{-3}$) in the $x$-$z$ coordinate plane, at time $t=600$ fm/c, immediately prior to the scission event for fission trajectory number 3.
The quadrupole deformation parameter is $\beta_{20} = 4.8$. Left middle and bottom panels: the corresponding proton $C_p$, and total $\sqrt{C_p C_n}$ localization functions, respectively. In the panels on the right the same plots are displayed, but the initial temperature is $T_{init} = 0.6$ MeV, and the temperature at scission $T_{sci} = 0.73$ MeV.
The scission event occurs at time $t=520$ fm/c, and the quadrupole deformation parameter is $\beta_{20} = 4.5$.
 }
 \label{fig:density_localization_250Cf_3}
\end{figure}

In the left top panel of Fig.~\ref{fig:density_localization_250Cf_3}, we plot the density profile of $^{250}$Cf (in units of fm$^{-3}$) in the $x$-$z$ coordinate plane, at time $t=600$ fm/c immediately prior to the scission event for fission trajectory number 3, for the case in which the initial point is at $T=0$, and the time evolution includes dynamical pairing correlations. The density profile at scission ($\beta_{20} = 4.8$) is characterized by the pronounced quadrupole and octupole deformation of the two large fragments, and an extended, low-density neck region. While the density does not exhibit any particular feature in the neck, the proton $C_p$ and total $\sqrt{C_p C_n}$ localization functions, shown in the left middle and bottom panels, respectively, reach peak values in the neck region that are much higher than typical nuclear matter values $\approx 0.5$ found in the bulk of the fragments. Here, the proton and neutron total localization functions are averaged over the spin: $C_{q}=(C_{q\uparrow}+C_{q\downarrow})/2$. Proton localization, in particular, reaches values close to 1, characteristic for $\alpha$-clusters.

\begin{figure}[tbh!]
\centering
\includegraphics[width=0.75\textwidth]{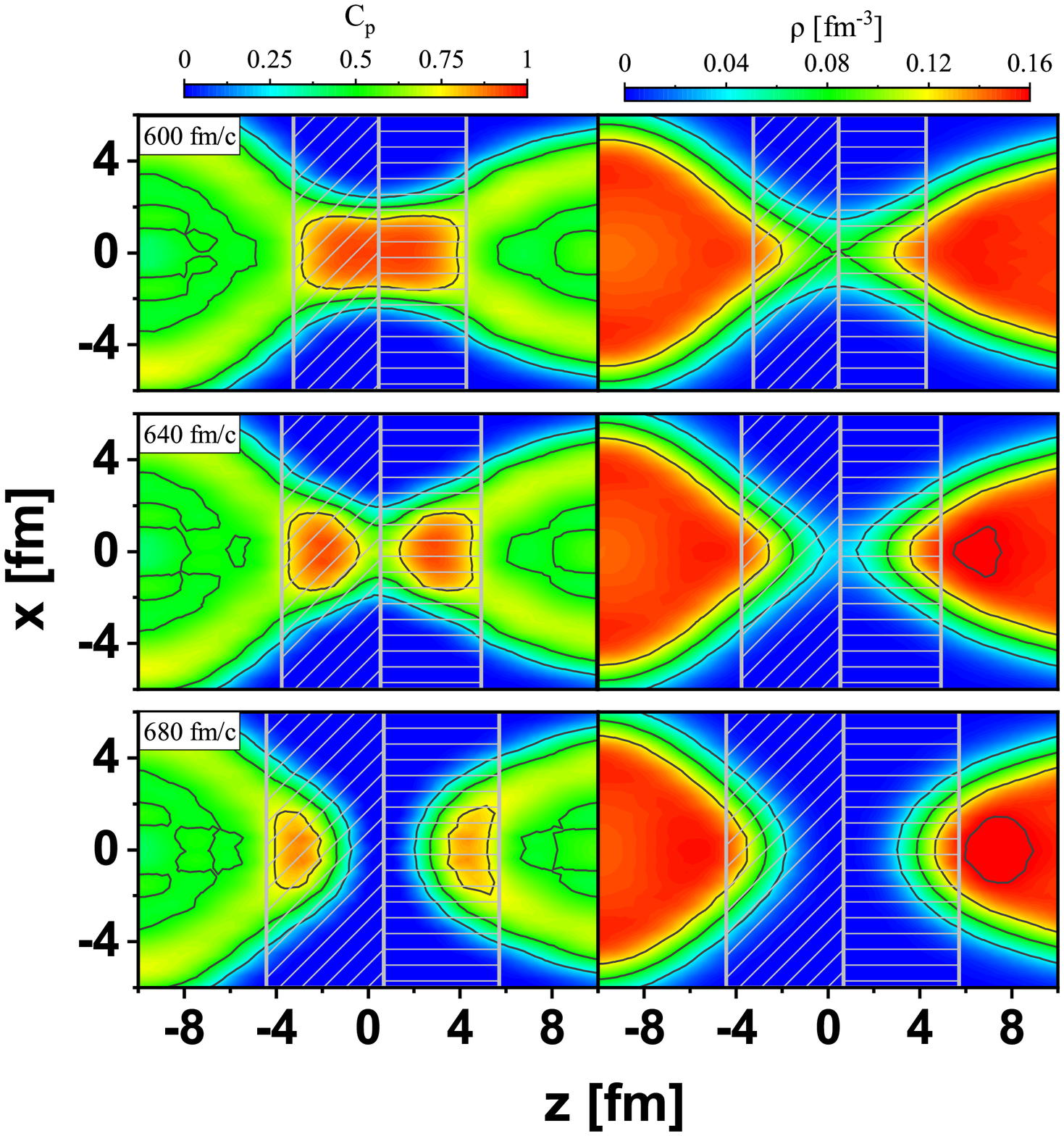}
 \caption{The proton localization function $C_p$ (left) and total density (right), at times: 600, 640, and 680 fm/c, for the fission trajectory number 3 of $^{250}$Cf. Starting from the point of lowest density along the $z$-axis, the shaded areas on the left and on the right denote regions that contains exactly two protons each.}
 \label{fig:neck_localization_250Cf_3}
\end{figure}

The scission event for trajectory number 3 is illustrated in Fig.~\ref{fig:neck_localization_250Cf_3}, where we display the proton localization function $C_p$ (left) and total density (right), at times immediately preceding scission (600 fm/c), at the moment when the fragments separate (640 fm/c), and immediately after (680 fm/c), when the separated fragments accelerate because of Coulomb repulsion.
Starting from the point of lowest density along the $z$-axis, the shaded areas on the left and on the right denote regions that contains exactly two protons each. The localization function clearly shows that the elongation of the neck region along the fission axis corresponds to two cluster containing two protons each. The number of neutrons in this region is almost double, the values of the corresponding localization function are somewhat lower and, therefore, we cannot uniquely identify $\alpha$-clusters. However, based on the argument of the much larger binding energy of $^{4}$He, the formation of $\alpha$-particles should be favored with respect to other light clusters, such as $^{3}$H and $^{6}$He.

\begin{figure}[tbh!]
\centering
\includegraphics[width=0.75\textwidth]{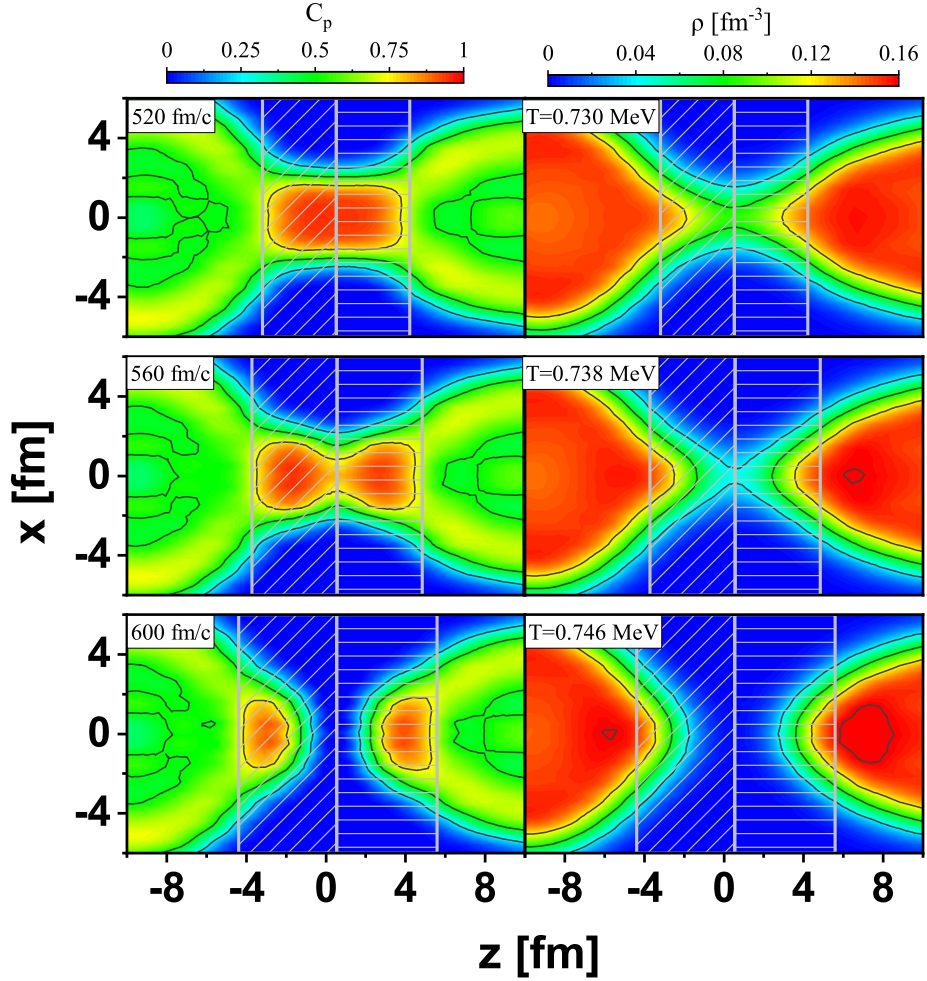}
 \caption{The proton localization function $C_p$ (left) and total density (right), at times: 520, 560, and 600 fm/c, for the fission trajectory number 3 of $^{250}$Cf. The initial temperature is $T_{init} = 0.6$ MeV, and the temperature at scission $T_{sci} = 0.73$ MeV. Starting from the point of lowest density along the $z$-axis, the shaded areas on the left and on the right denote regions that contains exactly two protons each.}
 \label{fig:neck_localization_250Cf_FinT_3}
\end{figure}

The results shown in the left panel of Fig.~\ref{fig:density_localization_250Cf_3} and in Fig.~\ref{fig:neck_localization_250Cf_3}, are very similar to those obtained for $^{240}$Pu in Ref.~\cite{Ren_22PRL}, and confirm that the time-scale of the formation of the neck, and the scission mechanism are governed by the dynamics of light clusters. In the right panel of Fig.~\ref{fig:density_localization_250Cf_3} and in Fig.~\ref{fig:neck_localization_250Cf_FinT_3}, we again display the density profiles and localization functions for trajectory number 3, but now for the case in which the initial state of the compound nucleus is at the temperature $T=0.6$ MeV, which  corresponds to thermal neutron-induced fission of $^{250}$Cf \cite{Brown_18NDS}. Except for a small difference in the elongation of the nucleus at scission, and a slightly shorter time it takes for the nucleus to reach scission, the increase in temperature ($T=0.73$ MeV just before scission) seems to have no significant effect on the formation of the clusters in the neck region.

\begin{figure}[tbh!]
\centering
\includegraphics[width=0.75\textwidth]{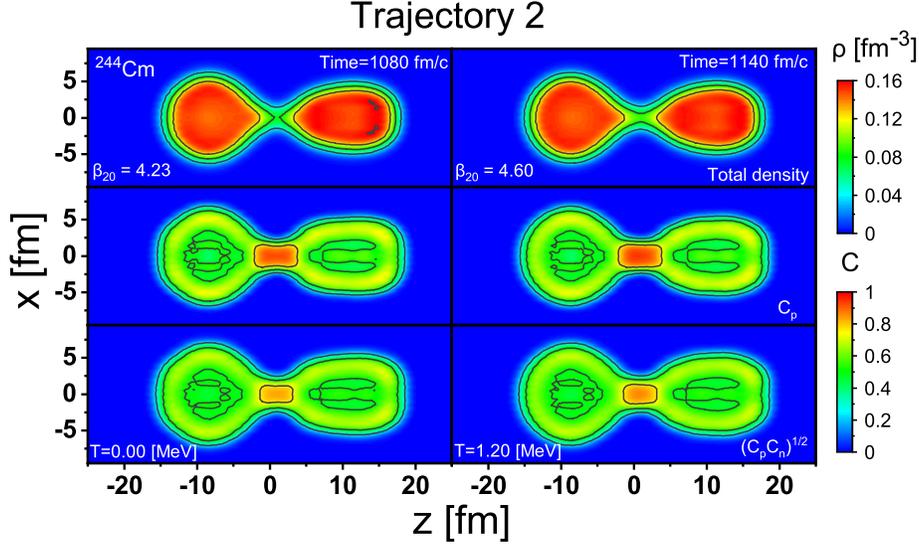}
\caption{Left top: density profile of $^{244}$Cm (color code in fm$^{-3}$) in the $x$-$z$ coordinate plane, at time $t=1080$ fm/c, immediately prior to the scission event for fission trajectory number 2. The quadrupole deformation parameter is $\beta_{20} = 4.23$. Left middle and bottom panels: the corresponding proton $C_p$ and total $\sqrt{C_p C_n}$ localization functions, respectively. In the panels on the right the same plots are displayed, but
the initial temperature is $T_{init} = 1.1$ MeV, and the temperature at scission $T_{sci} = 1.2$ MeV.
The scission event occurs at time $t=1140$ fm/c, and the quadrupole deformation parameter is $\beta_{20} = 4.6$.}
 \label{fig:density_localization_244Cm_2}
\end{figure}

\begin{figure}[tbh!]
\centering
\includegraphics[width=0.75\textwidth]{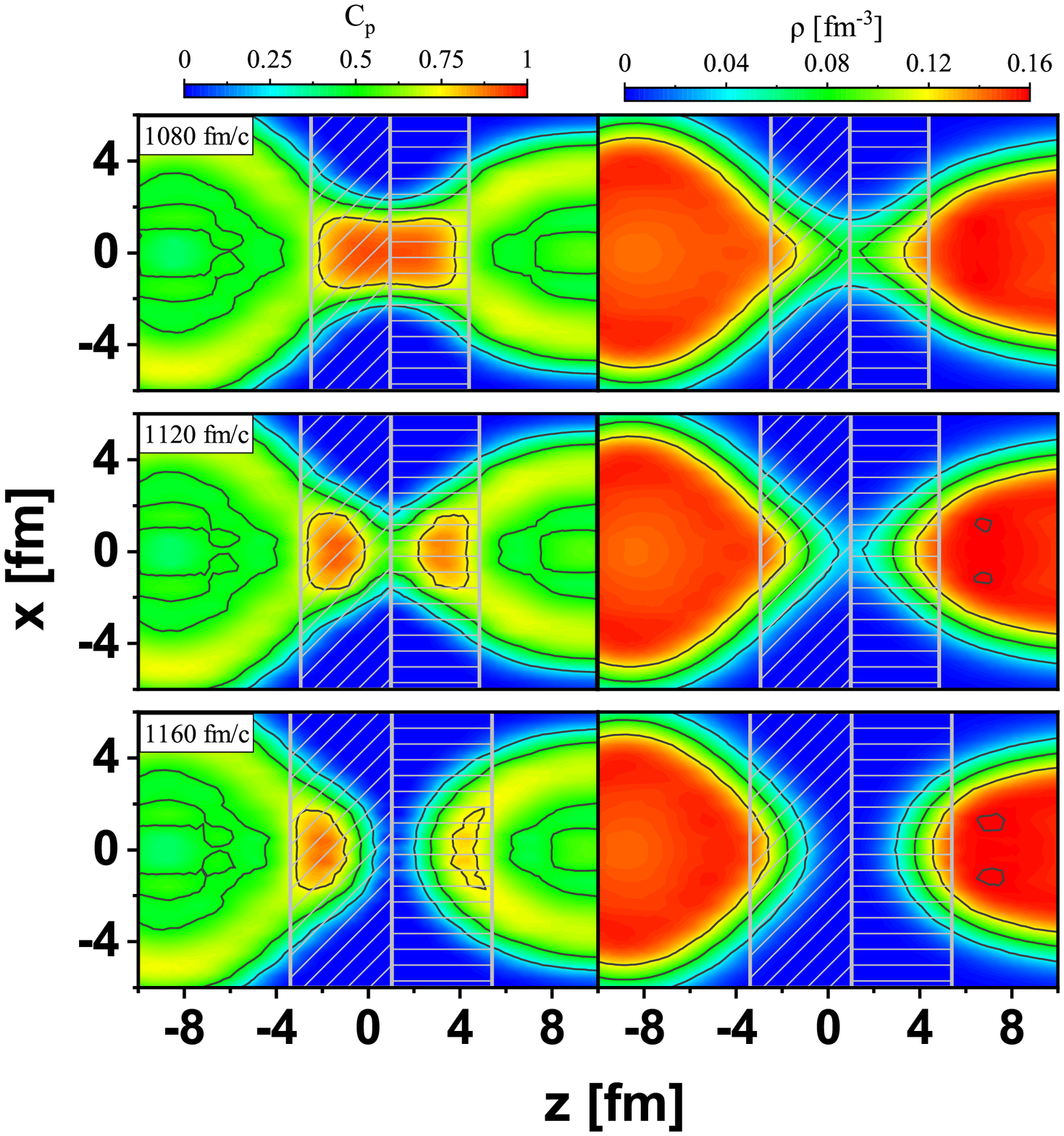}
 \caption{The proton localization function $C_p$ (left) and total density (right), at times: 1080, 1120, and 1160 fm/c, for the fission trajectory number 2 of $^{244}$Cm. Starting from the point of lowest density along the $z$-axis, the shaded areas on the left and on the right denote regions that contains exactly two protons each.}
 \label{fig:neck_localization_244Cm_2}
\end{figure}

\begin{figure}[tbh!]
\centering
\includegraphics[width=0.75\textwidth]{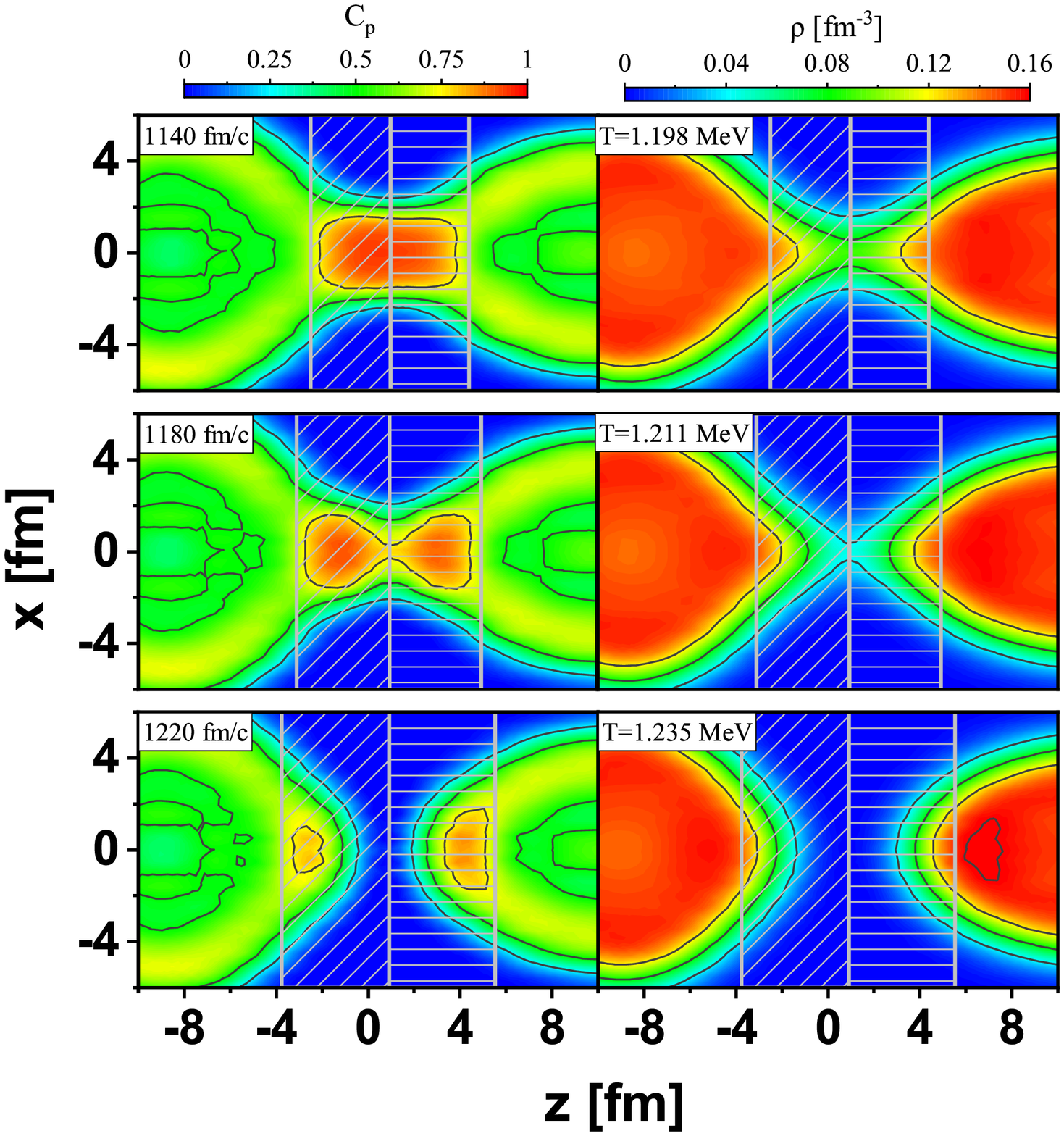}
 \caption{The proton localization function $C_p$ (left) and total density (right), at times: 1140, 1180, and 1220 fm/c, for the fission trajectory number 2 of $^{244}$Cm. The initial temperature is $T_{init} = 1.1$ MeV, and the temperature at scission $T_{sci} = 1.2$ MeV. Starting from the point of lowest density along the $z$-axis, the shaded areas on the left and on the right denote regions that contains exactly two protons each.}
 \label{fig:neck_localization_244Cm_FinT_2}
\end{figure}

In the second representative example, we have analyzed fission trajectory number 2 in $^{244}$Cm. In addition to the case with $T=0$ at the initial point (left panel of Fig.~\ref{fig:density_localization_244Cm_2} and Fig.~\ref{fig:neck_localization_244Cm_2}), the results obtained for $T_{init} = 1.1$ MeV are shown in
right panel of Fig.~\ref{fig:density_localization_244Cm_2} and in Fig.~\ref{fig:neck_localization_244Cm_FinT_2}. In the latter case the initial temperature corresponds to an average experimental excitation energy of the compound nucleus of 23 MeV for multinucleon transfer-induced fission \cite{Ramos2018_PRC97-054612}. Even though this is the highest excitation energy among the examples considered in the present study, and the temperature at scission reaches $T_{sci} = 1.2$ MeV, it appears that this temperature is not high enough to prevent the formation of light clusters in the low-density neck region. Also in this case, the difference between the results for the density profiles and localization functions at scission, obtained with $T_{init} = 0$ and $T_{init} = 1.1$ MeV, is not significant. In fact, we have verified that the pronounced nucleon localization and consequently the formation of light clusters in the low-density neck region at times immediately preceding scission, is a robust result for all fission trajectories considered in the four nuclei: $^{240}$Pu, $^{234}$U, $^{244}$Cm, and $^{250}$Cf.

\section{Summary}\label{sec_summ}

A microscopic finite-temperature model based on time-dependent nuclear density functional theory (TDDFT), has been applied to analyze the saddle-to-scission dynamics of induced fission of $^{240}$Pu, $^{234}$U, $^{244}$Cm, and $^{250}$Cf. In a recent study \cite{Zhao2019_PRC99-054613}, we have investigated the induced fission dynamics of these nuclei in the finite temperature TDGCM+GOA framework. Here, in addition to the standard zero-temperature TDDFT approach in which pairing correlations are treated dynamically with the time-dependent BCS approximation \cite{ren20LCS,ren20O16,Ren_22PRC}, we have developed a finite-temperature TDDFT formalism that allows to follow the changes in temperature along fission trajectories. Even though the present implementation of the self-consistent method does not include the dynamical treatment of pairing correlations at finite temperature, it is nevertheless very useful for a realistic description of fission dynamics in cases in which the excitation energy of the compound system corresponds to temperatures that are well above the pairing phase transition, that is, for which pairing correlations vanish.

For each of the four illustrative nuclei, we have considered three characteristic initial points beyond the outer barrier, at energies approximately 1 MeV below the equilibrium minimum. Given the initial single-nucleon wave functions and occupation probabilities, the zero-temperature and finite-temperature TDDFT models propagate the nucleons independently toward scission and beyond. We have compared self-consistent fission trajectories that are obtained starting the time-evolution at zero temperature and treating pairing correlations dynamically, with those that are computed when the initial temperature corresponds to the experimental excitation energy of the fissioning system. Since the trajectories represent the final phase of the fission process, very similar results are obtained at $T=0$ and finite temperature, both for the paths that basically follow the route of steepest descent in the collective space of quadrupole and octupole deformations, and for the lengths of the time interval from the initial point of a trajectory to the corresponding scission point.

Very interesting results have been obtained with the finite-temperature TDDFT analysis of saddle-to-scission dissipative dynamics.
Starting from the initial values, the non-equlibrium generalization of the chemical potential and temperature are adjusted at each step so that the particle number and total energy, respectively, are conserved along a TDDFT trajectory. This results in an isentropic fission path, that is, the local entropy remains constant along the TDDFT trajectory. The corresponding increase in temperature between the initial point and scission is of the order of 10\% to 20\%. By partitioning the total energy into various kinetic and excitation energy contributions, it has been shown that: (i) only a smaller part of the potential energy difference between the initial and scission points is converted into collective flow energy; (ii) the dissipated energy is at least a factor $2 - 4$ larger than the prescission kinetic energy. Quantitative results have been obtained for the deformation energies of the fragments at scission and, therefore, for the total intrinsic excitation energy at scission. For the examples that have been considered in the present study, the initial temperatures range from $0.6$ Mev for thermal neutron-induced fission of $^{250}$Cf, to $1.1$ MeV for multinucleon transfer-induced fission of $^{244}$Cm with an average experimental excitation energy of 23 MeV. The prescission kinetic energies are calculated in the interval between $4$ and $9$ MeV, depending on the specific nucleus and fission trajectory, while the dissipated energy ranges between $12$ and $27$ MeV.

In the second part of this work, the finite-temperature TDDFT has been applied to the dynamics of neck formation and rupture. In a recent study of fission dynamics of $^{240}$Pu \cite{Ren_22PRL}, we have shown that the time-scale of formation of a low-density neck between the nascent fragments coincides with the assembly of two $\alpha$-like clusters. The length of the neck corresponds to the spatial extension of the two clusters, and at scission the neck ruptures between the clusters, which separate because of Coulomb repulsion and are absorbed by the two heavy fragments. Since these results were obtained for a single illustrative case of $^{240}$Pu, to verify the universality of the proposed scission mechanism, here we have performed additional calculation of fission trajectories in the four actinide nuclei, both at zero and finite temperatures, as described above. The new results have confirmed those obtained in Ref.~\cite{Ren_22PRL}, that is, in all cases at times immediately preceding scission a region of high nucleon localization is formed between the emerging fragments. The localization function for protons reaches values close to one, characteristic for $\alpha$-particles and, by integrating over the one-body density, we have shown that the neck region contains four protons, while the number of neutrons is almost twice as large. Although at the mean-field level one cannot distinguish between different clusters, because of the much larger binding energy of $^{4}$He,  $\alpha$-clusters should dominate over $^{3}$H and $^{6}$He.

Another reason for applying the self-consistent finite-temperature TDDFT formalism to neck dynamics, is the possibility to follow the increase in temperature along fission trajectories, especially in the neck region at times preceding scission. This is important because, in general, cluster formation will be suppressed by the heating of low-density matter between fragments. However, for realistic initial  temperatures that correspond to experimental excitation energies, and an increase of 10\% to 20\% between the initial and scission points, no significant difference in the localization functions at scission has been observed with respect to paths that started at zero temperature. For final temperatures between 0.7 MeV ($^{250}$Cf) and 1.3 MeV ($^{244}$Cm), the energy dissipated along the fission paths is simply not large enough to prevent the formation of clusters, favored by the appearance of a low-density region between the two heavy fragments.

Finally, in the present analysis axial symmetry has been assumed, that is, the starting points of TDDFT trajectories have been determined in the space of quadrupole and octupole collective parameters $\beta_{20}$ and $\beta_{30}$, that characterize axially symmetric deformation energy surfaces. Consequently, the light clusters appearing in the neck region, are always absorbed by the heavy fragments at the moment of scission, inducing strongly damped dipole oscillations along the fission axis. To observe ternary fission events in which one of the clusters is not absorbed by the corresponding heavy fragment, axial symmetry needs to be broken. We started considering such initial points already in our previous study \cite{Ren_22PRL}, but so far have not been able to induce a fission process in which more than two fragments are produced. Ternary fission thus remains an intriguing topic for future theoretical studies in the TDDFT framework.

\begin{acknowledgments}
This work has been supported in part by the High-end Foreign Experts Plan of China,
National Key R\&D Program of China (Contracts No. 2018YFA0404400),
the National Natural Science Foundation of China (Grants No. 12070131001, 11875075, 11935003, 11975031, and 12141501),
the High-performance Computing Platform of Peking University,
the QuantiXLie Centre of Excellence, a project co-financed by the Croatian Government and European Union through the European Regional Development Fund - the Competitiveness and Cohesion Operational Programme (KK.01.1.1.01.0004),
 and the Croatian Science Foundation under the project Uncertainty quantification within the nuclear energy density framework (IP-2018-01-5987).
J. Z. acknowledges support by the National Natural Science Foundation of China under Grants No. 12005107 and No. 11790325.
\end{acknowledgments}

\bigskip

%
\end{document}